# Distributed Contingency Analysis over Wide Area Network among Dispatch Centers

Zhengwei Ren,
Ying Chen,
Shaowei Huang
Dept. of Electrical Engineering
Tsinghua University
Beijing, China

Shuang Sheng
Dept. NARI Technology Development
Limited Company
Beijing, China

Huiping Zheng,
Xinyuan Liu
Electric Power Research Institute
Shanxi Electric Power Grid
Taiyuan, China

*Abstract*—Traditionally, a regional dispatch center uses the equivalent method to deal with external grids, which fails to reflect the interactions among regions. This paper proposes a distributed N-1 contingency analysis (DCA) solution, where dispatch centers join a coordinated computation using their private data and computing resources. A distributed screening method is presented to determine the Critical Contingency Set (DCCS) in DCA. Then, the distributed power flow is formulated as a set of boundary equations, which is solved by a Jacobi-Free Newton-GMRES (JFNG) method. During solving the distributed power flow, only boundary conditions are exchanged. Acceleration techniques are also introduced, including reusing preconditioners and optimal resource scheduling during parallel processing of multiple contingencies. The proposed method is implemented on a real EMS platform, where tests using the Southwest Regional Grid of China are carried out to validate its feasibility.

## I. INTRODUCTION

Normally, a regional dispatch center only has real-time measurements and parameters of its territory grid [1], while the external grids are represented by proper equivalents during contingency analysis (CA). However, the equivalent method may not adequately reflect the interaction among regional grids. Currently, as complicated data integration procedures are involved, a joint CA of an interconnected power system with multiple regions can only be conducted by a superior dispatch center. When regional dispatch centers want to study risky interactions of their local controls, they may lack real-time data as well as accurate analysis tools. Therefore, in this work, the distributed contingency analysis method is studied, which takes advantages of coordinated computation among dispatch centers.

Several methods have been proposed to model the influences from external grids during the security analysis of regional power system. In [2] and [3], the Thevinan and Ward equivalence were used to represent the external grids, with periodical corrections. In [4], a master-slave-splitting method was used to improve the accuracy of external grids equivalents, which depends on solving the global power flow by alternative iterations.

It should be noted that with advances in communication technologies, distributed computations among dispatch centers become feasible and reliable, which tends to be more accurate than isolated and equivalent based simulations. Following this idea, models and algorithms have been developed for coordinated analyses among dispatch centers. In [5], JFNG algorithm is proposed to solve the power flow with higher convergence. In [6], distributed power flow with JFNG algorithm is carried out, which shows good convergence even for multiple regions. In [7] and [8], distributed power flows were used in contingency analysis of interconnected power systems, where all regional power grids perform power flow simultaneously and exchange boundary information through Wide Area Network (WAN). In [9], the allocation of network loss is considered by coordinating the active power output of slack buses and giving revisions on the injection power.

During CA, it is necessary to choose a Critical Contingency Set (CCS) to reduce the number of contingencies that need full AC power flow. Distribution factor was widely applied to fast determine the CCS in [10], [11], which is simple but not so reliable. In [12], the eigenvalue sensitivity method was proposed for screening the CCS resulting in large disturbances. Further study in [13] introduced the first-order eigenvalue-sensitivity method, which depends on the estimation of system eigenvalues. Due to sensitivity analysis requires the overall power grid, above methods are not suitable for distributed contingency analyses among dispatch centers, where parameters of the whole power system may be unavailable.

In this paper, a DCA method is proposed and tested for a real interconnected power system. Utilizing computation and data resources in dispatch centers, a flexible architecture of DCA is introduced. The main workflow of the DCA is also elaborated. Then, a distributed CCS screening (DCCS) is developed for enhancing the efficiency of DCA, which iteratively searches critical lines according to their electrical distances to boundary buses. Moreover, a distributed power flow model is established, which is solved by a JFNG method

This work is supported in part by the National Natural Science Foundation of China (51677100), China State Grid Corp Science and Technology Project (SGSXDKY-DWKJ2015-001) and in part by the NSFC under Grants 51277104 and 51321005.

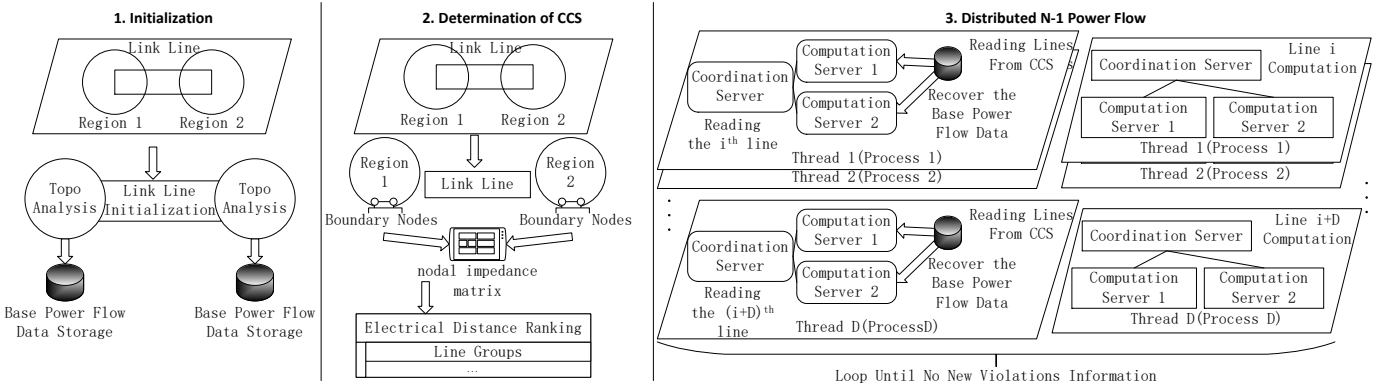

Figure 1. Diagram of DCA

and incorporated into the parallel contingency analyses. Finally, the proposed DCA is implemented on an EMS platform, where case studies are carried out for a real interconnected power grid.

The rest of this paper is organized as follow. In section II, the architecture of distributed contingency analysis is proposed and in section III the distributed CCS screening method is introduced. Section IV establishes the power flow model and section V provides the implementation of the architecture and the optimized scheduling for paralleled computation. Section VI validates the distributed power flow results and illustrates the efficiency of acceleration methods as well as the computation on DCCS. Section VII makes conclusions.

## II. ARCHITECTURE OF DCA

In DCA, the regional dispatch centers are coordinated with their data and resources. By communication through WAN, a unified N-1 contingency analysis results can be achieved. To accelerate the computation speed, the multithreading and multiprocessing techniques are used in this architecture.

### A. DCA over Wide Area Network

In distributed CA, regions exchange information through WAN. Logically there are mainly two parts which are the coordination servers and the computation servers. The computation servers are in charge of the analysis of the regional grid. The coordination server is responsible for the receiving, processing, and delivery of the boundary data and messages. The distributed CA is initiated through coordination server. All computation servers work simultaneously during the distributed CA.

The coordination server gives orders such as initialization, the start of computation, data transmit confirmation, stop of computation, etc. The communication networks of distributed CA are based on WAN. Message heads are designed to distinguish data and messages from different threads and handshake protocol ensures the reliability of communication.

### B. Workflow of DCA

DCA mainly consists of the following steps: 1) The initialization of information and power flow data; 2) The determination of CCS; 3) The distributed N-1 power flow with paralleled technique. Procedures are shown in figure 1.

1) The initialization of information and power flow data. The coordination server initializes the N-1 CA and sends initialization information to the computation servers. Message heads and parameters are confirmed during initialization. For computation servers, base power flow is carried out basing on the topology analysis of the grid structure.

2) Determination of CCS. The base power flow computation can provide the nodal impedance matrix for each regional grid. The nodes are ranked according to electrical distances and lines are grouped with the nodes they connect to. A criterion is adopted to determine the number of lines in the CCS which is also the stopping criterion.

3) Distributed N-1 power flow. The computation servers will firstly recover the power flow data such as grid structures. Then, coordination server will read the line names from the CCS. After that, the computation servers will modify the power flow data according to the line names, and distributed power flow (parallelization in algorithm level) is performed. To accelerate the computation speed, multithreading and multiprocessing techniques (parallelization in data level) are implemented.

A stopping criterion is used to stop the DCA. The procedure 3) is looped and meanwhile violation information is collected and compared by groups. If there are no new differences between the violation information of the adjacent groups, the loop will be stopped.

## III. DISTRIBUTED CCS SCREENING

The purpose of CCS screening is to reduce the computation burden and time cost. In the real EMS system, the CCS is screened according to the line voltage levels with DC power flow. For different voltage levels, the number of lines to be analyzed are given empirically. Thus a more theoretic method is necessary.

In distributed CA, the cross-regional influences are passed on through link lines. Specifically, an internal line contingency changes the grid operation status which will affect the boundary injection power. If the changing of boundary injection power is great enough, the limit violation might occur in the other region. Therefore, the nearer the lines are to the boundary nodes, the more possible that the cross-regional influences may happen.

More attentions should be paid to the lines adjacent to the boundary nodes. The electrical distance is used to evaluate the relationship between two nodes in the power grid and equals to the impedance between two nodes. The electrical distance can be obtained from the nodal impedance matrix. The nearer the distance is, the closer the relationship is.

The computation of electrical distance for different regions is carried out by different computation servers separately with the following steps:

1) Different computation server establishes the nodal admittance matrix for its regional power grid;
2) The nodal impedance matrix is obtained by inverting the admittance matrix;
3) The electrical distances of internal nodes to boundary nodes are obtained from admittance matrix and ranked in ascending order respectively.

The computation of the inverse of admittance matrix is only once for the whole DCA, so the computation is not of great complexity. Different computation servers compute simultaneously to get their list of nodes. The lines are grouped according to the nodes which they connect. For each computation server, there will be a CCS.

## IV. DISTRIBUTED POWER FLOW FOR CONTINGENCY ANALYSIS

The DCA is based on the N-1 distributed power flow. In this section, a distributed power flow model considering network loss allocation is adopted. The interconnected power system is divided into two regional grids and a link line partition. Then boundary equations are established on the distributed power flow model and JFNG method [5] is used to solve the boundary equations.

### A. System Partitioning

Figure 2 shows the partitioning of a 2-region interconnected power system.

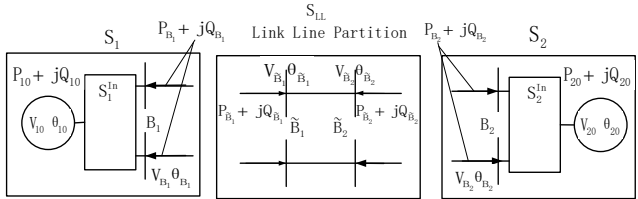

Figure 2. Partitioning of Interconnected Power System

Let $i = \{1,2\}$. $P_{B_i}$ and $Q_{B_i}$ represent the boundary injection power of partition $S_i$; $P_{\tilde{B}_i}$ and $Q_{\tilde{B}_i}$ represent the boundary injection power of link line partition. $V_{B_i}$ and $\theta_{B_i}$ are the voltage amplitudes and angles of boundary nodes for $S_i$, while $V_{\tilde{B}_i}$ and $\theta_{\tilde{B}_i}$ are for link line partition. for each region, there is one slack bus, which is denoted by $V_{i0}$, $\theta_{i0}$, $P_{i0}$ and $Q_{i0}$. The slack bus is often chosen from a PV bus. Among various slack buses, a dominant slack bus is chosen to set the base value of angle for the whole interconnected power system.

### B. Boundary Equations

The type of boundary nodes is PQ node. Let $\boldsymbol{U}_{B_1} = V_{B_1} \angle \theta_{B_1}$, $\boldsymbol{U}_{B_2} = V_{B_2} \angle \theta_{B_2}$. Then for $S_1$ and $S_2$, there are:

$$\boldsymbol{U}_{B_1} = f_B^1(\theta_{10}, P_{B_1}, Q_{B_1}) \quad (1)$$

$$\boldsymbol{U}_{B_2} = f_B^2(\theta_{20}, P_{B_2}, Q_{B_2}). \quad (2)$$

For link line partition, $\boldsymbol{U}_{\tilde{B}_1} = V_{\tilde{B}_1} \angle \theta_{\tilde{B}_1}$, $\boldsymbol{U}_{\tilde{B}_2} = V_{\tilde{B}_2} \angle \theta_{\tilde{B}_2}$; $\boldsymbol{U}_{\tilde{B}_1} = \boldsymbol{U}_{B_1}$ and $\boldsymbol{U}_{\tilde{B}_2} = \boldsymbol{U}_{B_2}$. There are:

$$\begin{cases} P_{\tilde{B}_1} = f_{P_1}^{Lk}(\boldsymbol{U}_{\tilde{B}_1}, \boldsymbol{U}_{\tilde{B}_2}) \\ P_{\tilde{B}_2} = f_{P_2}^{Lk}(\boldsymbol{U}_{\tilde{B}_1}, \boldsymbol{U}_{\tilde{B}_2}) \\ Q_{\tilde{B}_1} = f_{Q_1}^{Lk}(\boldsymbol{U}_{\tilde{B}_1}, \boldsymbol{U}_{\tilde{B}_2}) \\ Q_{\tilde{B}_2} = f_{Q_2}^{Lk}(\boldsymbol{U}_{\tilde{B}_1}, \boldsymbol{U}_{\tilde{B}_2}) \end{cases} \quad (3)$$

For non-dominant slack buses, take region 1 for example, the original reactive power output $P_{10}$ is given by the PV bus. For different $\theta_{10}$, there is different active power $\tilde{P}_{10}$, which is:

$$\tilde{P}_{10} = f_{In}^1(\theta_{10}, P_{B_1}, Q_{B_1}). \quad (4)$$

$f_B^1, f_B^2, f_{P_1}^{Lk}, f_{P_2}^{Lk}, f_{Q_1}^{Lk}, f_{Q_2}^{Lk}$ and $f_{In}^1$ all refer to the power flow equations. The boundary equations are given as:

$$\begin{cases} \Delta P_{B_i} = P_{B_i} + P_{\tilde{B}_i} \\ \Delta Q_{B_i} = Q_{B_i} + Q_{\tilde{B}_i}, i = 1,2. \\ \Delta P_{10} = P_{10} - \tilde{P}_{10} \end{cases} \quad (5)$$

Therefore, the necessary and sufficient conditions for the converged power flow of the whole interconnected power system are that $\Delta P_{B_i}, \Delta Q_{B_i}$ and $\Delta P_{10}$ should all equal to zero.

### C. Algorithm

JFNG(m)[5] is a modified Newton-GMRES method with high convergence. A preconditioner is used to approach the inverse of Jacobi matrix through rank-one updating during iterations. The initial value for the preconditioner is usually an identity matrix. The more similar the preconditioner is to the inverse of Jacobi matrix, the higher the convergence of the JFNG is. As the explicit form of Jacobi matrix does not need to be obtained, for equations with great complexity, the JFNG simplifies the process of computing the inverse of Jacobi matrix.

## V. SYSTEM IMPLEMENTATION AND OPTIMIZATION

### A. D5000 Simulation Platform

The proposed method in this paper is tested in the real power system. D5000 is a platform used for operation and dispatch in the State Grid Corporation of China (SGCC) [14]. It integrates several functions such as state estimation, power flow, and contingency analysis, etc.

1) Communication System: D5000 platform provides the WAN Event Service for exchanging data and messages. The coordination and computation servers can send and receive information by calling the interfaces. The average communication time is about 0.4 millisecond.

2) Software System: D5000 platform is installed on Linux operation system. The platform has its power flow application but can only compute the local region. New applications can be developed and expanded on D5000 platform.

3) Hardware System: High-performance servers are used in the test system. Each server has 8 CPUs. The main frequency of CPU is 3.0GHz, produced by ADM. The internal memory is 16GB.

## B. Accelerate Convergence of Distributed Power Flow

In JFNG, two methods are frequently used to improve the convergence of the solution of boundary equations. The higher the convergence is, the fewer the numbers of iterations are, and less time will be cost.

### 1) Reuse of the initial value for boundary equations

For the paralleled architecture, when a line contingency computation is converged, final values of independent variables for boundary equations can be obtained. If the line contingencies are near in electrical distance, the final values may be similar. Thus the initial values of next round computation can reuse the results of the previous round.

### 2) Reuse of preconditioner

For contingencies that are similar, the Jacobi matrix of power flow should also be similar. For JFNG algorithm, the more the preconditioner approaches the inverse of Jacobi matrix, the higher the convergence is. To speed up the process, for each round of computation, the preconditioner reuses its latest value of previous round as the initial value.

## C. Optimized Schedule for Parallelism

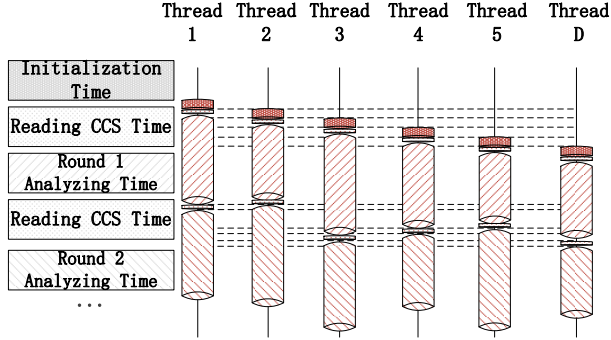

Figure 3. Paralleled Computation for DCA

The paralleled technique can greatly lower the total time cost with adequate scheduling. In DCA, two levels of parallelization are incorporated, which are algorithm level and data level. The former refers to the distributed power flow using JFNG algorithm, while the latter refers to evaluating the multiple contingencies by multi-thread processing. While parallelizing contingency analyses, the next contingency should be read from CCS and analyzed right after the previous one. Figure 3 shows the paralleled computation for DCA. The reading of CCS is serial in order to avoid memory conflict.

Each round of analyzing time includes the computation time, communication time and coordination time. The communication through WAN is almost synchronous, and the computation and coordination time are proportional to the numbers of iterations. The total time $T$ depends on the thread with the longest time, which can be expressed as:

$$T = \max\{t_{d1}, t_{d2}, \ldots, t_{dD}\}. \quad (6)$$

For $k = \{1, 2, \ldots, D\}$, $t_{dk}$ is the time cost for thread $k$, while $D$ is the number of threads.

## VI. TEST RESULTS

### A. Test Systems

The XN system is used in this paper, which is a real system from the southwest regional grid of China. It consists of two provinces, the partitioning of the XN system is based on the two provinces. Table I lists the details of the XN system. There are four link lines which are 275 to 206 and 333 to 501. Both of them are double circuit lines.

TABLE I. XN POWER GRID

| System Name | Bus Numbers | Branch Numbers | Boundary Node Number |
|---|---|---|---|
| XN Area 1 | 1039 | 1517 | 275, 333 |
| XN Area 2 | 213 | 388 | 206, 501 |

### B. Validation of Distributed Power Flow

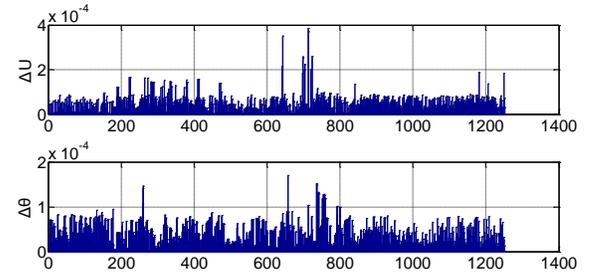

Figure 4. Absolute Residuals of Centralized and Distributed Power Flow

Figure 4 gives the absolute residuals of centralized and distributed power flow results of the XN test system. The maximum residual is less than $4 \times 10^{-4}$, which is within the acceptance of accuracy.

### C. Contingency Analysis Based on CCS

The distributed N-1 power flow is computed in groups. Table II shows the computation order. Since parallelism is adopted, multiple lines can be computed at the same time, so the CA can be finished in short time.

TABLE II. CONTINGENCY ANALYSIS BASED ON DISTRIBUTED CCS

| From Node | XN Area | Electrical Distance | To Node | Violation Information |
|---|---|---|---|---|
| 333 | 1 | 0.00187 | 202*; 365; 469*; 264* | $P_{max}$ violation in Area 2 |
| 335 | 1 | 0.00241 | 1216; 876; 883 | None Violation |
| 989 | 1 | 0.00241 | 335 | None Violation: Stop |
| 501 | 2 | 0.01548 | 545; 483* | $P_{max}$ violation in Area2 |
| 503 | 2 | 0.01620 | 966*; 1000; 1038 | None Violation |
| 1044 | 2 | 0.01620 | 503* | None Violation: Stop |

\* Double-circuit lines

It can be seen from Table II, the actual number of lines which need full AC distributed power flow is very small.

### D. Convergence Analysis with Acceleration Methods

Figure 5 shows the total numbers of iterations for each line contingency in Table II. The results show that the reuse of initial value and preconditioner can greatly enhance the computation convergence.

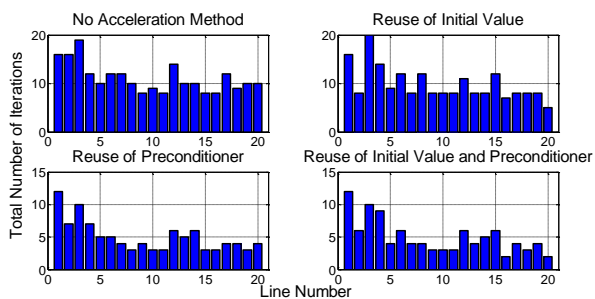

Figure 5. Total Numbers of Iterations

*E. Choosing of Thread Number*

Equation (6) shows that the total time $T$ is determined by the thread with the longest time cost. Limited by computation resources on the D5000 platform, when $D$ grows, $T$ will first decrease then increase. For a given system, $D$ is decided using practical tests with a constant total number of line contingencies, as shown in figure 6. The specific value of $D$ may vary according to the real system environment.

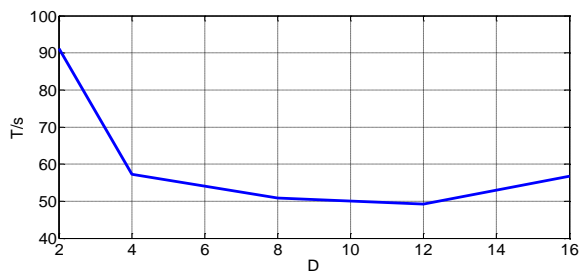

Figure 6. Relationship between $T$ and $D$

In figure 6, for each value of $D$, $T$ is chosen as the maximum value among replicated tests. It can be seen that when $D = 12$, $T$ is at its minimum value while the given system is at its best performance.

Tests of branch outages in XN system on D5000 platform using parallelization can accomplish the distributed N-1 contingency analysis within 60 seconds. There are totally 1905 line contingencies that are analyzed, including the full distributed power flow computation for 20 lines listed in Table II.

## VII. CONCLUSION

In this study, a DCA method is proposed to consider the regional interactions in N-1 contingency analysis based on distributed power flow model, which avoids complicated data integrations of the regional power grids. The DCCS screening is proposed to relief the computation burden. To accelerate the computation, multithreading and multiprocessing technique are adopted in coordination servers and computation servers respectively with optimized scheduling. The reuse of initial value and preconditioner of JFNG are proved to lower the numbers of iterations. The DCA method is verified on a real EMS system by finishing the contingency analysis within 60 seconds. In the future, possible acceleration methods will be explored to enhance the efficiency of the DCA in WAN.